\documentclass[a4paper,11pt]{article}

\usepackage{geometry} 
\usepackage[utf8]{inputenc} 
\usepackage[T1]{fontenc} 
\usepackage{lmodern}
\usepackage[british]{babel}
\usepackage{graphicx}
\graphicspath{{figs/}} 
\usepackage[pdfusetitle]{hyperref}
\usepackage{xspace} 

\usepackage{microtype} 
\usepackage[margin=15mm,format=hang]{caption} 
\usepackage{xfrac} 
\makeatletter
\g@addto@macro\@floatboxreset\centering
\makeatother

\usepackage[intlimits]{amsmath} 
\usepackage{amssymb} 
\usepackage{bm} 
\usepackage{bbm} 
\numberwithin{equation}{section} 
\allowdisplaybreaks 

\usepackage{cite} 
\usepackage[capitalise]{cleveref} 
\creflabelformat{equation}{\textup{#2#1#3}} 

\usepackage{braket} 
\usepackage{siunitx} 
\usepackage{suffix} 

\renewcommand{\Re}{\operatorname{Re}}
\renewcommand{\Im}{\operatorname{Im}}
\newcommand{\BigO}[1]{\ensuremath{\mathcal{O}\left(#1\right)}}

\newcommand{\Lagrangian}{\ensuremath{\mathcal{L}}\xspace}
\newcommand{\hc}{\text{h.c.}}

\newcommand{\V}[1]{\ensuremath{V_{#1}^{}}} 
\WithSuffix\newcommand\V*[1]{\ensuremath{V_{#1}^*}}

\newcommand{\program}[1]{\texttt{#1}}
\newcommand{\flavio}{\program{flavio}\xspace}
\newcommand{\smelli}{\program{smelli}\xspace}
\newcommand{\wilson}{\program{wilson}\xspace}
\newcommand{\HighPT}{\program{HighPT}\xspace}
\newcommand{\QuantumNumbers}[3]{(\mathbf{#1}, \mathbf{#2}, #3)}
\newcommand{\nuDIS}{\ensuremath{\nu\text{DIS}}\xspace}

\title{A \texorpdfstring{$\bm{\nu}$}{nu} window onto leptoquarks?}
\author{Matthew Kirk, Shohei Okawa, Keyun Wu}

\begin{document}

\begin{center}
\vspace*{1cm}
{\LARGE\bfseries
\makeatletter\@title\makeatother
}
\\[0.8 cm]
\textsc{
Matthew Kirk, Shohei Okawa, Keyun Wu
}
\\[0.5 cm]
{\small
Institut de Ciències del Cosmos (ICCUB) and Departament de Física Quàntica i Astrofísica (FQA), Universitat de Barcelona (UB), Spain
\\[0.2cm]
E-mail:
\texttt{mjkirk@icc.ub.edu},
\texttt{okawa@icc.ub.edu},
\texttt{keyunwu@fqa.ub.edu}
}
\end{center}

\vskip0.5cm

\renewcommand{\abstractname}{\Large\bfseries Abstract}

\begin{abstract}
Upcoming neutrino telescopes promise a new window onto the interactions of neutrinos with matter at ultrahigh energies ($E_\nu =$ \qtyrange{1e7}{1e10}{\GeV}), and the possibility to detect deviations from the Standard Model predictions.
In this paper, we update previous predictions for the enhancement of the neutrino-nucleon cross-section for motivated leptoquark models and show the latest neutrino physics bound, as well as analyse the latest LHC pair production and Drell-Yan data, and flavour constraints (some of which were previously missed).
We find that, despite the next generation of neutrino experiments probing the highest energies, they will not be enough to be competitive with collider searches.
\end{abstract}

{\small \tableofcontents}

\section{Introduction}
\label{sec:intro}

There is a new generation of neutrino experiments on the horizon that will probe neutrino interactions with matter at some of the highest energies ever measured, up to \qty{1e10}{\GeV}.
A recent proposal \cite{Esteban:2022uuw} has examined how, assuming any ultrahigh energy (UHE) neutrino experiment reaches some minimal detector requirements, the data can be combined to make a measurement of the neutrino-nucleon scattering cross-section at the highest energies (similar measurements from particular experiments have been discussed in \cite{Denton:2020jft,Valera:2022ylt}).

As the centre-of-mass energies for the neutrino interactions can be as high as \qty{100}{\TeV}, these measurements provide a window into physics far beyond the reach of current colliders.
We expect the Standard Model (SM) to break down at some energy scale above the $\sqrt{s} \sim \BigO{\qty{1}{\TeV}}$ at which it has been tested thus far, and so there is the clear potential to search for beyond the Standard Model (BSM) physics -- much work has been done here previously analysing the potential for BSM searches (see \cite{Goncalves:2022uwy,GarciaSoto:2022vlw,Huang:2022ebg,Anchordoqui:2001cg,Marfatia:2015hva} for examples).
In \cite{Esteban:2022uuw} they briefly touched upon this, showing that leptoquarks (LQs) or extra dimension theories could alter the predicted SM cross-section by large amounts.
For LQs, this enhancement arises through resonant s-channel production in neutrino-nucleon scattering events which is an unavoidable feature of any LQ model with non-zero couplings, as well as through gluon scattering that is again a necessary feature since LQs have colour charge.

In this work we take this idea forward, choosing two well motivated LQ models ($R_2$ and $S_1$) to test against this new search strategy.
The $R_2$ LQ is motivated by the ongoing hints of lepton-flavour universality violating new physics in $b \to c \ell \nu$ decays (in which context it has been well studied, see e.g.\ \cite{Datta:2017aue,Chen:2017hir,Kamali:2018fhr,Angelescu:2018tyl,Popov:2019tyc,Iguro:2020keo,Angelescu:2021lln,Crivellin:2022mff,Fedele:2022iib}), and has been previously investigated using the results of lower energy neutrino measurements as a constraint \cite{Becirevic:2018uab}.
Our other model of study, the $S_1$, was more recently studied for its potential to be found using cosmic rather than collider experiments, this time using the energy and angular distributions of the events at specific upcoming experiments \cite{Huang:2021mki}.
In particular, that work used a gap in the collider search results that allowed very light LQs to still remain viable, which enhanced the discovery potential from neutrino deep inelastic scattering (\nuDIS).

In this article, we will summarise the upcoming neutrino experiments in \cref{sec:neutrinos}, discussing future precision and our calculation of the BSM enhancement.
In \cref{sec:other_constraints} we will discuss the other limits on LQ models.
Firstly from LHC searches for pair production (\cref{sub:direct_lhc}) and high $p_T$ Drell-Yan measurements (\cref{sub:highPT}), where we will find that the latest measurements have massively increased the mass limits on LQ models.
Next we look to flavour in \cref{sub:flavour}, where will discuss the $R_{D^{(*)}}$ anomaly and perform an updated fit, and provide a bound from LQ mediated tau decay that was missed in the earlier work of \cite{Huang:2021mki}.
Finally we use precise electroweak precision observables (EWPO) calculations and a new result for electric dipole moments to find strong bounds on particular areas of parameter space in \cref{sub:ewpo,sub:EDM}.
With all these results in hand, we then combine all this data in \cref{sec:combination} and show that even in the very best case for our LQ models and optimistic scenario for future neutrino data, it will be hard for the neutrino experiments to exceed the current bounds.

\section{Leptoquark models}

In this section we give the details of our chosen LQ models, and their interaction Lagrangians and coupling structures.

\subsection{$R_2$}

The $R_2$ is a scalar LQ with the $\QuantumNumbers{SU(3)_c}{SU(2)_L}{U(1)_Y}$ quantum numbers $\QuantumNumbers{3}{2}{\sfrac{7}{6}}$.
The $R_2$ can be a potential explanation of the $R_{D^{(*)}}$ anomalies but only with large couplings (see the later discussion in \cref{sub:bclnu}), which makes it a good potential candidate for the new search analysis.

For ease of comparison, we use the same notation and minimal coupling structure as \cite{Becirevic:2018uab} (which is sufficient to explain the $R(D^{(*)})$ anomalies):
\begin{equation}
\Lagrangian = g_R^{ij} \bar{Q}_i R_2 e_j - g_L^{ij} \bar{u}_i R_2 \epsilon L_j + \hc
\end{equation}
where $Q$ is the left-handed SM quark doublet, $L$ is the left-handed SM lepton doublet, $u$ is the right-handed SM up-type quark, and $e$ is the SM right-handed charged lepton, and with 
\begin{equation}
g_L = 
\begin{pmatrix}
0 & 0 & 0 \\
0 & 0 & g_L^{c\tau} \\
0 & 0 & 0
\end{pmatrix},
\qquad
g_R = 
\begin{pmatrix}
0 & 0 & 0 \\
0 & 0 & 0 \\
0 & 0 & g_R^{b\tau}
\end{pmatrix} \,.
\end{equation}
After electroweak (EW) symmetry breaking, the LQ couplings have the form
\begin{equation}
\Lagrangian = (V \cdot g_R)_{ij} \bar{u}_i P_R \ell_j R_2^{5/3} + (g_R)_{ij}  \bar{d}_i P_R \ell_j R_2^{2/3} + (g_L)_{ij} \bar{u}_i P_L \nu_j R_2^{2/3} - (g_L)_{ij} \bar{u}_i P_L \ell_j R_2^{5/3} + \hc
\end{equation}
where $\ell$ represents a charged lepton, and we have assumed the CKM rotations are entirely within the up sector.
For the purposes of our neutrino cross-section calculation, we neglect the CKM entirely, since for the $R_2$ there are no extra CKM suppressed but parton distribution function (PDF) enhanced neutrino-quark interactions lost by this assumption (see also the discussion in the next section for the $S_1$).

\subsection{$S_1$}

Our other LQ choice is the $S_1$, with quantum numbers $\QuantumNumbers{3}{1}{\sfrac{-1}{3}}$.
This was chosen for detailed study in \cite{Huang:2021mki} as it had property of being able to tune the couplings such that it primarily decayed to tau leptons plus light quarks, a decay mode that was unsearched for at the LHC at that time.
As we will see later however, this gap in the search has now been filled, rendering the LHC constraints significantly stronger.

Our Lagrangian (again we follow the notation of \cite{Huang:2021mki}, which e.g.\ differs by a complex conjugation in the definition of $S_1$ from \cite{Dorsner:2016wpm}) is 
\begin{equation}
\Lagrangian = y_{LL}^{ij} \overline{Q^c_i} \epsilon L_j S_1^* + y_{RR}^{ij} \overline{u^c_i} e_j S_1^* + \hc
\end{equation}
with
\begin{equation}
y_{LL} = 
\begin{pmatrix}
0 & 0 & 0 \\
0 & 0 & y_{LL}^{s\tau} \\
0 & 0 & 0
\end{pmatrix},
\qquad
y_{RR} = 
\begin{pmatrix}
0 & 0 & 0 \\
0 & 0 & y_{RR}^{c\tau} \\
0 & 0 & 0
\end{pmatrix} \,.
\end{equation}
After EW symmetry breaking, the LQ couplings have the form
\begin{equation}
\Lagrangian = -y_{LL}^{ij} \overline{d^c_i} P_L \nu_j S_1^{1/3} + (V^* \cdot y_{LL})^{ij} \overline{u^c_i} P_L \ell_j S_1^{1/3} + y_{RR}^{ij} \overline{u^c_i} P_R \ell_j S_1^{1/3} + \hc
\end{equation}
where we have again assumed CKM rotations in the up sector, to match the choice made in \cite{Huang:2021mki}.
One might consider whether the choice of whether to use the up or down-basis for the CKM rotations could lead to a difference in the cross-section, since in principle it is clear that this could open up new CKM suppressed but PDF enhanced channels.
However it turns out that, in the important region of the integral, the PDF enhancement is smaller than the CKM suppression (which will be at least $V_{us}^2 \sim 1/20$), and so the total cross-section is relatively insensitive to the choice we have made.

\section{Neutrino interactions}
\label{sec:neutrinos}

Neutrinos are some of the most mysterious and least understood particles in the SM, due to the difficulty of observing their interactions. 
However natural cosmic processes lead to a flux of UHE neutrinos through the Earth, which although as yet undetected, hold the potential to deliver data on interactions at even higher energies than our current generation of particle colliders -- the centre of mass energy of a neutrino-proton collision at ultrahigh energies ($E_\nu =$ \qtyrange{1e7}{1e10}{\GeV}) is roughly $\sqrt{s} \sim$ \qtyrange{4}{140}{\TeV}, which at the high end exceeds even the energy of any currently proposed future collider!
Discussion within the neutrino community about these UHE measurements dates back many years \cite{Kusenko:2001gj,Hooper:2002yq,Hussain:2006wg,Borriello:2007cs}, but only recently has it become plausible for the measurements to be made.
It is clear then that this data should be analysed for its potential to shed light on the expected breakdown of the SM.

\subsection{Neutrino experiments and BSM searches}
\label{sub:nu_searches}

Previous work in the literature has already used current and projected upcoming neutrino data to search for deviations from the SM.
In \cite{Becirevic:2018uab}, they studied the $R_2$ LQ, as motivated by the $R_{D^{(*)}}$ anomalies, and the sensitivity of the IceCube experiment to this model, including a forward looking extrapolation to future larger data sets, and their results form the first main comparison for this present work.
More recently, the authors of \cite{Huang:2021mki} performed a systematic investigation of new physics (NP) models that would alter the signal at the upcoming tau neutrino experiments GRAND, POEMMA and Trinity.
At that time, they found that the $S_1$ LQ was the least constrained by collider searches, and was thus the candidate with the most potential in the neutrino experiments, and we therefore make our second comparison to their results.
Another study of several radiative neutrino mass LQ models using potential resonant effects in IceCube and non-standard neutrino interactions (NSI) was done in \cite{Babu:2022fje}, finding no potential for IceCube to outperform LHC searches.
Finally, the work of \cite{Esteban:2022uuw} studied how the observations from all upcoming UHE neutrino experiments could be combined to make a measurement of the \nuDIS cross-section, without any knowledge of the currently unknown cosmic neutrino flux.%
\footnote{In that work they only assume that the flux can be modelled by a simple power law, which is well justified since the low statistics and small energy range studied mean this is a good approximation regardless of the true spectral shape, and they demonstrate that the precise spectral index has no effect on their results.}
In particular we note that the results of \cite{Esteban:2022uuw} ignore potential neutrino regeneration while in passage through the Earth, which is a model dependent affect, since they show that these are subdominant in the SM at the energies and precision levels achievable in the near future.
We have studied the differential neutrino cross-section in our BSM scenarios, and confirmed that they very closely match the SM one at the neutrino energies we consider, and hence regeneration continues to be subdominant (see \cref{app:nu_regeneration} for more details).

This means that their forecast measurement of the \nuDIS cross-section can be easily applied to any NP model, and the sensitivity compared to other bounds in a simple way with minimal calculation required.
In contrast, the works of \cite{Becirevic:2018uab} and \cite{Huang:2021mki} carefully calculate the full propagation of neutrinos in their respective BSM models, which gives a technically more accurate result but requires more in-depth calculations on a model by model basis.

In \cite{Esteban:2022uuw}, they show that neutrino-nucleon cross-section can be measured to ${}^{+\qty{65}{\percent}}_{-\qty{30}{\percent}}$ precision with a small number of UHE neutrino events, or even $\pm\qty{15}{\percent}$ in an optimistic scenario of larger statistics and more experiments (very similar results were found in \cite{Denton:2020jft,Valera:2022ylt}).
In that work, they are agnostic about the flavour sensitivity of the future neutrino experiments that will contribute data, although this will affect the precise nature of the cross-section that can be measured.
Future experiments like IceCube-Gen2 \cite{IceCube-Gen2:2020qha}, ARA \cite{Allison:2011wk} or RNO-G \cite{RNO-G:2020rmc} will be able to measure a mix of all lepton flavours, and so this data will give a measurement of the flavour averaged \nuDIS cross-section (assuming a flavour universal cosmic flux, which the current data supports \cite{IceCube:2015gsk,IceCube:2018pgc,IceCube:2020fpi}, although we note that even in the SM this incoming flux universality will be broken at the detectors by an \BigO{\qty{1}{\percent}} boost to the tau neutrino flux from UHE electron and muon neutrinos \cite{Soto:2021vdc}).
On the contrary, tau neutrino telescope experiments (for example GRAND \cite{GRAND:2018iaj,Kotera:2021hbp}, TAMBO \cite{Romero-Wolf:2020pzh}, Trinity \cite{2019BAAS...51g..67O} or POEMMA \cite{POEMMA:2020ykm}) are sensitive only to the decay products from tau neutrinos, and so data exclusively from these would result in a measurement of the flavour specific $\nu_\tau$ interaction cross-section (which we will denote as $\sigma_\tau$).
Such a flavour specific measurement is very beneficial when considering the potential power to discover NP that is correspondingly flavour specific.
Consider the total cross-section ratio for a generic (i.e.\ non flavour specific) model:
\begin{equation}
\frac{\sigma}{\sigma^{SM}} = 1 + \frac{\sum_i \sigma^{BSM}_i}{\sum_i \sigma^{SM}_i} = 1 + \frac{\sum_i \sigma^{BSM}_i}{3 \sigma^{SM}_\tau} \,.
\end{equation}
where in the second equality we have used the fact that, to a very good approximation, the SM \nuDIS cross-section is equal for each flavour.
If in addition we are able to measure the cross-section ratio for a specific flavour (taking tau as an example for obvious reasons), we can write that theoretical prediction as
\begin{equation}
\frac{\sigma_\tau}{\sigma^{SM}_\tau} = 1 + \frac{\sigma^{BSM}_\tau}{\sigma^{SM}_\tau} \,.
\end{equation}
Finally we see that, for a flavour specific NP model, where $\sum_i \sigma^{BSM}_i = \sigma^{BSM}_\tau$, the expected enhancement relative to the SM will be three times larger for the flavour specific measurement than for the all flavour result, or equivalently that the sensitivity to flavour specific NP is three times greater.

We calculate the LQ enhancement for our two different models, and compare the future sensitivity bounds from \cite{Esteban:2022uuw} (for both flavour averaged and tau specific cross-section measurements) to previous attempts detailed above to use neutrino experiments to search for LQs.

\subsection{Cross-section calculation}
\label{sub:nuDIS_xsec}

\begin{figure}
\includegraphics[viewport=140 600 455 770, clip=true, scale=0.65]{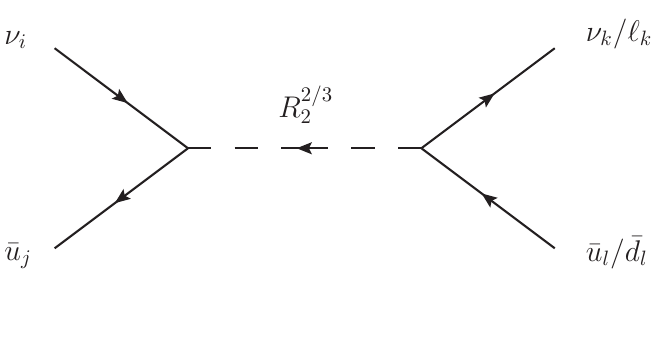}
\includegraphics[viewport=78 560 585 770, clip=true, scale=0.65]{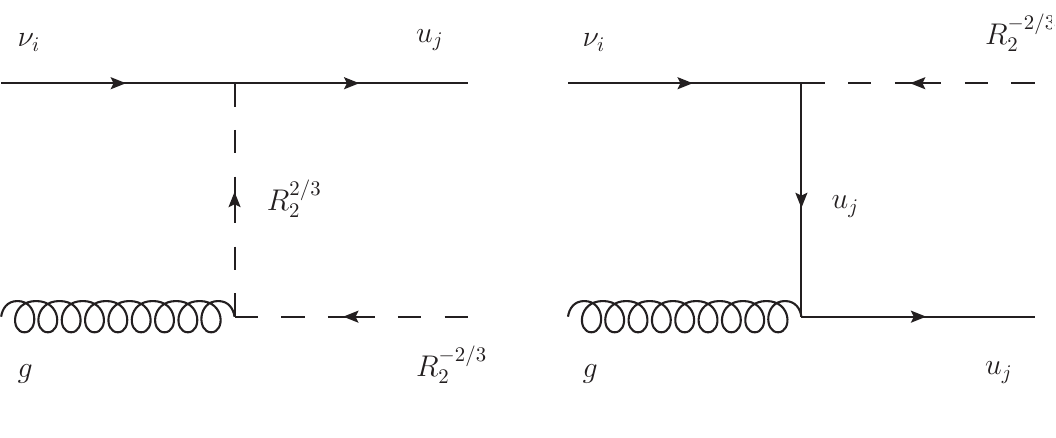}
\caption{Feynman diagrams showing the important \nuDIS matrix elements for the $R_2$ LQ. The equivalent diagrams for $S_1$ can be found by replacing $u \leftrightarrow \bar d$, $\bar u \leftrightarrow d$.}
\label{fig:R2_diagrams}
\end{figure}

\begin{figure}
\includegraphics[width=0.96\textwidth]{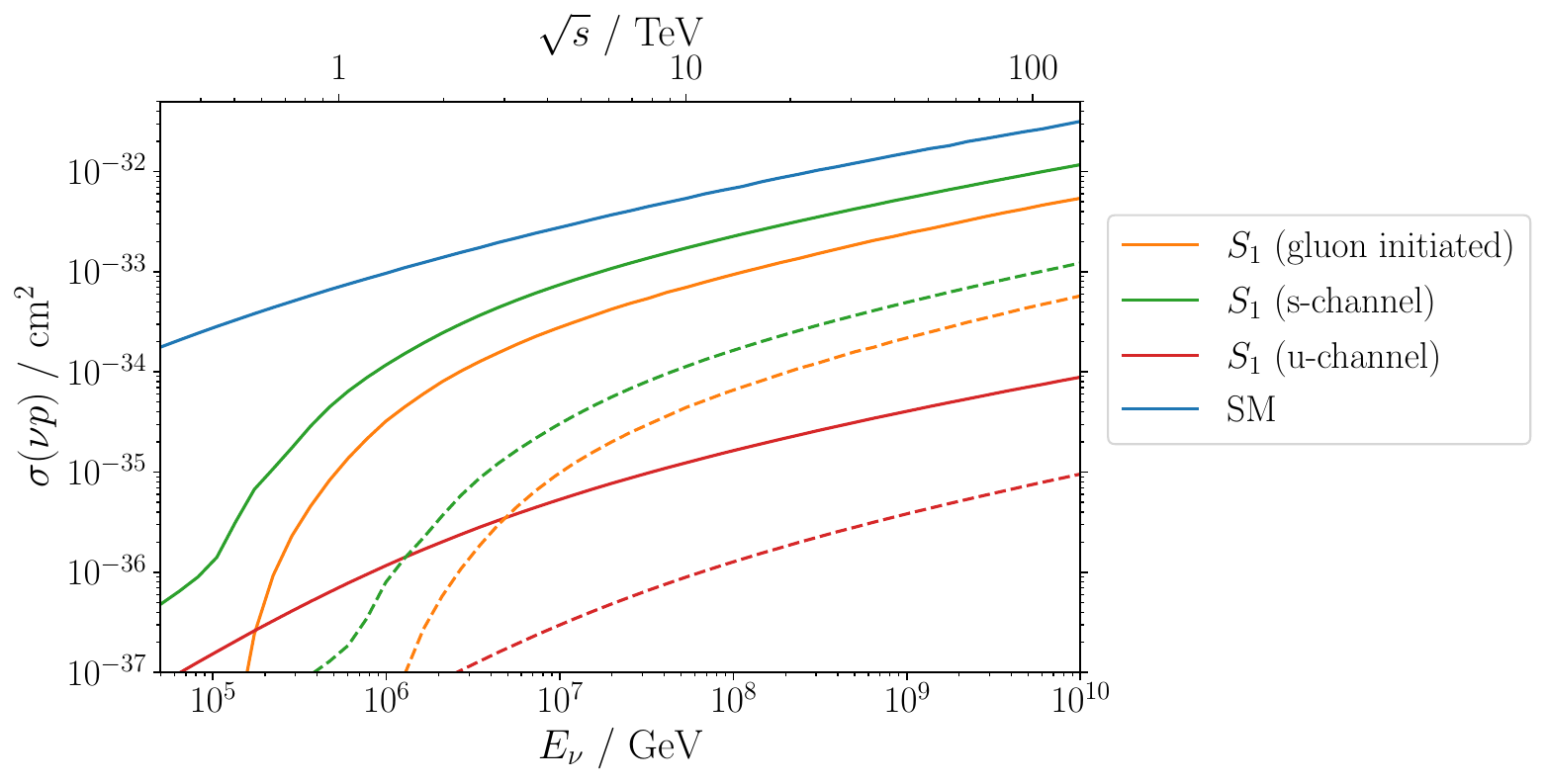}   
\caption{Comparing the different subprocesses involved in \nuDIS scattering for an $S_1$ LQ with a mass of \qty{400}{\GeV} (solid lines) or \qty{1}{\TeV} (dashed lines) with couplings $y_{LL}^{s\tau} = y_{RR}^{c\tau} = 1$.
Very similar results are found for the $R_2$ LQ (see also Figure 9 in \cite{Becirevic:2018uab}).}
\label{fig:process_comparison}
\end{figure}

The total cross-section for neutrino scattering from a proton can be written as 
\begin{equation}
\sigma( \nu p ) = \sum_q \int_{x_\text{min}}^{1} \int_{y_\text{min}}^{y_\text{max}} q(x, Q^2) \frac{\braket{|M (\nu q \to )|^2}}{16 \pi \hat{s}} + \int_{x_\text{min}}^{1} \int_{y_\text{min}}^{y_\text{max}} g(x, Q^2) \frac{\braket{|M (\nu g \to )|^2}}{16 \pi \hat{s}}
\end{equation}
where $q$ and $g$ are the PDFs for quarks and gluons at some particular momentum fraction $x$ and momentum transfer $Q^2$, and the $\braket{|M|^2}$ are the spin-averaged matrix element squared for the different possible processes (see \cref{fig:R2_diagrams}).
In addition to the more obvious enhancement possible from LQs due to s-channel resonant diagrams, it was shown in \cite{Becirevic:2018uab} that there is also a large contribution to the neutrino-nucleon cross-section from gluon initiated diagrams, particularly when the LQ primarily couples to heavy quarks, as is the case for our chosen models (and more generally for LQs in common NP models), since the gluon PDF grows rapidly at smaller $x$.
This can be seen for the $S_1$ LQ from \cref{fig:process_comparison}, where we see that for centre-of-mass energies above the LQ mass, the gluon initiated and s-channel cross-sections rapidly increase as the LQ becomes kinematically accessible.

In our integration, for the quark initiated processes we set the lower limits of the $x$ integral to \num{1e-9}, which is the specified lower limit of our chosen PDF set (see later discussion) beyond which the fitted PDF data is extrapolated,%
\footnote{We have checked that this does not have a significant effect on the result.}
while for the gluon initiated processes there is a kinematic limit of $x_\text{min} = (m_q + m_\text{LQ})^2 / (2 m_p E_\nu)$. 
For the $y$ kinematic variable, the limits are
\begin{equation}
\frac{\hat{s} - m_q^2 - \sqrt{\lambda(\hat{s}, m_q^2, 0)}}{2\hat{s}} < y < \frac{\hat{s} - m_q^2 + \sqrt{\lambda(\hat{s}, m_q^2, 0)}}{2\hat{s}}
\end{equation}
for $\nu$-$q$ scattering, and 
\begin{equation}
\frac{\hat{s} - m_q^2 - M_\text{LQ}^2 - \sqrt{\lambda(\hat{s}, M_\text{LQ}^2, m_q^2)}}{2\hat{s}} < y < \frac{\hat{s} - m_q^2 - M_\text{LQ}^2 + \sqrt{\lambda(\hat{s}, M_\text{LQ}^2, m_q^2)}}{2\hat{s}}
\end{equation}
for $\nu$-$g$ scattering, where the Källén function is defined as $\lambda (a,b,c) = a^2 + b^2 + c^2 - 2ab - 2ac - 2bc$.
In our notation, for a process $\nu \psi_2 \to \psi_3 \psi_4$ the Mandelstam kinematic invariants are defined as 
\begin{equation}
\hat{s} = (p_\nu + p_2)^2 \,, \quad \hat{t} = (p_\nu - p_3)^2 \,, \quad \hat{u} = (p_\nu - p_4)^2
\end{equation}
with the neutrino and other initial particle momenta incoming, and the final particle momenta outgoing.

We give below the matrix elements for the important resonant and gluon initiated terms, with the rest to be found in \cref{app:nuDIS_matrix_elements}.
(In all cases, we neglect quarks masses, which is a good approximation for our couplings choices, as well as CKM rotations as previously discussed, and calculate only to leading order in the perturbative expansion.)

\paragraph{$R_2$}

\begin{align}
\braket{|M(\nu_i \bar{u}_j \to \nu_k \bar{u}_l)|^2} &= 
\begin{aligned}[t]
&\frac{\hat{s}^2}{2} \frac{|g_{L}^{ji}|^2 |g_{L}^{lk}|^2}{(\hat{s} - M_{R_2}^2)^2 + M_{R_2}^2 \Gamma_{R_2}^2}
\\
&+ \delta_{ik} \delta_{jl} \left( 
- \frac{8 \alpha \pi}{3 c_W^2} \frac{\hat{s}^2 (\hat{s}-M_{R_2}^2)}{((\hat{s} - M_{R_2}^2)^2 + M_{R_2}^2 \Gamma_{R_2}^2)(\hat{t}-M_Z^2)} \Re(g_{L}^{ji} g_{L}^{lk*})
\right.
\\
 &\qquad \qquad \, \left.
 + \frac{2 \alpha^2 \pi^2 (16 \hat{s}^2 s_W^4 + (3-4 s_W^2)^2 \hat{u}^2)}{9 c_W^4 s_W^4 (\hat{t} - M_Z^2)^2}
 \right)
\end{aligned}
\\
\braket{|M(\nu_i \bar{u}_j \to \ell_k \bar{d}_l)|^2} &= 
\begin{aligned}[t]
&\frac{\hat{s}^2}{2} \frac{|g_{L}^{ji}|^2 |g_{R}^{kl}|^2}{(\hat{s} - M_{R_2}^2)^2 + M_{R_2}^2 \Gamma_{R_2}^2}
\\
&+ \delta_{ik} \delta_{jl} \left( + \frac{8 \alpha^2 \pi^2 \hat{u}^2}{s_W^4 (\hat{t} - M_W^2)^2}
 \right)
\end{aligned}
\\
\braket{|M(\nu_i g \to R_2^{-2/3} u_j)|^2} &=
\begin{aligned}[t]
-\frac{2 \alpha_s \pi \hat{s} (2 M_{R_2}^4 - 2 M_{R_2}^2 (\hat{s} + \hat{t}) + (\hat{s} + \hat{t})^2) |g_{L}^{ji}|^2}{\hat{t} (\hat{s} + \hat{t})^2}
\end{aligned}
\end{align}

%
%

\paragraph{$S_1$}

\begin{align}
\braket{|M(\nu_i d_j \to \nu_k d_l)|^2} &= 
\begin{aligned}[t]
&\frac{\hat{s}^2 |y_{LL}^{ji}|^2 |y_{LL}^{lk}|^2}{2 ((\hat{s} - M_{S_1}^2)^2 + M_{S_1}^2 \Gamma_{S_1}^2)}
\\
&+ \delta_{ik} \delta_{jl} \left( \frac{2 \alpha \pi}{3 c_W^2 s_W^2} \frac{\hat{s}^2 (\hat{s} - M_{S_1}^2) (3 - 2 s_W^2)}{((\hat{s} - M_{S_1}^2)^2 + M_{S_1}^2 \Gamma_{S_1}^2) (\hat{t} - M_Z^2)} \Re(y_{LL}^{lk *} y_{LL}^{ji}) \right.
\\
&\qquad \qquad \, \left. + \frac{2 \alpha^2 \pi^2}{9 c_W^4 s_W^4} \frac{\hat{s}^2 (3 - 2 s_W^2)^2 + 4 \hat{u}^2 s_W^4}{(\hat{t} - M_Z^2)^2}\right)
\end{aligned}
\\
\braket{|M(\nu_i d_j \to \ell_k u_l)|^2} &=
\begin{aligned}[t]
&\frac{\hat{s}^2 |y_{LL}^{ji}|^2 (|y_{LL}^{lk}|^2 + |y_{RR}^{lk}|^2)}{2 ((\hat{s} - M_{S_1}^2)^2 + M_{S_1}^2 \Gamma_{S_1}^2)}
\\
&+ \delta_{ik} \delta_{jl} \left( \frac{4 \alpha \pi}{s_W^2} \frac{\hat{s}^2 (\hat{s} - M_{S_1}^2)}{((\hat{s} - M_{S_1}^2)^2 + M_{S_1}^2 \Gamma_{S_1}^2) (\hat{t} - M_W^2)} \Re(y_{LL}^{lk *} y_{LL}^{ji}) \right.
\\
&\qquad \qquad \,  \left. + \frac{8 \alpha^2 \pi^2}{s_W^4} \frac{\hat{s}^2}{(\hat{t} - M_W^2)^2}\right)
\end{aligned}
\\
\braket{|M(\nu_i g \to S_1^{-1/3} \bar{d}_j)|^2} &=
\begin{aligned}[t]
- \frac{2 \alpha_s \pi \hat{s} (2 M_{S_1}^4 - 2 M_{S_1}^2 (\hat{s} + \hat{t}) + (\hat{s} + \hat{t})^2) |y_{LL}^{ji}|^2}{\hat{t} (\hat{s} + \hat{t})^2}
\end{aligned}
\end{align}

\begin{figure}
\includegraphics[width=\textwidth]{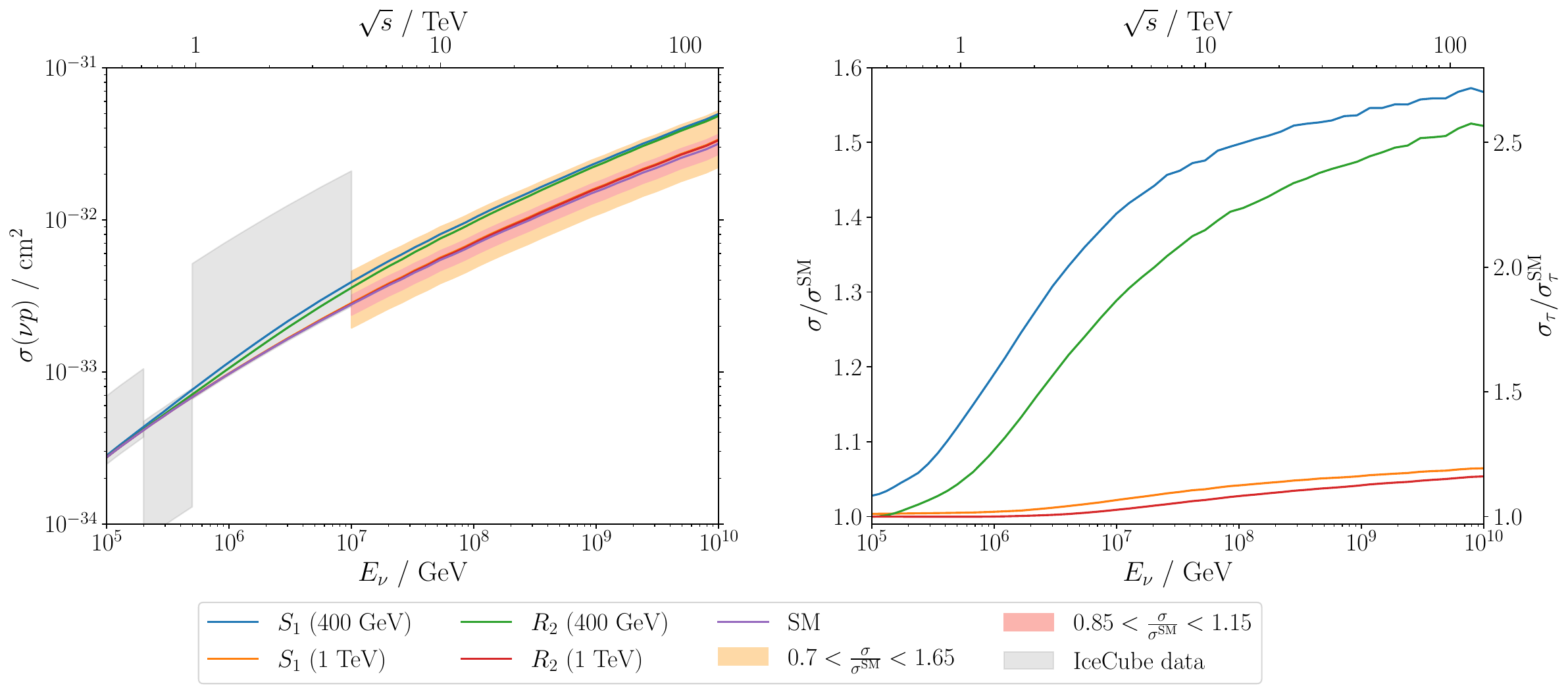}
\caption{Neutrino-proton cross-section for the SM and various LQ models (left), and the corresponding ratios (right) which are the basis for our neutrino bounds.
The IceCube data is taken from \cite{IceCube:2020rnc}, and the coloured shaded bands shows the expected measurement precision from upcoming neutrino experiments as calculated in \cite{Esteban:2022uuw}.
For the $R_2$, we fix $g_L^{c\tau} = g_R^{b\tau} / (0.75i) = 1$, while for the $S_1$ we fix $y_{LL}^{s\tau} = y_{RR}^{c\tau} = 1$.}
\label{fig:nuDIS_xsec}
\end{figure}

In \cref{fig:nuDIS_xsec} we show the neutrino-proton scattering cross-section as a function of the incoming neutrino energy, for the SM as well as our different LQ scenarios.%
\footnote{Readers may note that we do not reproduce the extremely large LQ enhancement seen in the first published version of \cite{Esteban:2022uuw}. Following discussion with the authors, we confirmed that our calculation were correct and identified a small mistake in their work, which has since been  corrected in the latest version.}
We see that the ratio of the BSM cross-section compared to the SM grows with energy, and so for our constraints we will use the ratio at the highest neutrino energy.
Furthermore however, we see that this cross-section enhancement drops off rapidly at larger LQ masses, which as we will see in \cref{sec:combination} limits the reach of neutrino experiments.

\subsubsection{Discussion of theoretical uncertainty}
\label{sub:nu_pdf_errors}

The theoretical calculation of the \nuDIS cross-section requires knowledge of nucleon constituents when probed at extremely high energies, corresponding to very small parton momentum fraction $x$.
As an example, consider that the s-channel LQ scattering has a resonance at values close to $M_\text{LQ}^2 / (2 m_N E_\nu)$, which for a \qty{1}{\TeV} LQ and the neutrino energies we are considering can be as small as \num{5e-5}.
In recent work by the CTEQ collaboration \cite{Xie:2023suk}, they studied the PDF uncertainties associated with the SM cross-section at energies up to \qty{1e12}{\GeV}, and showed (see FIG.\ 13 in their article) that currently the uncertainty is of order \qty{10}{\percent} at those energies.
We have studied the effects of different PDF choices on the LQ cross-sections, and find similar results.

There is also the question of which is the appropriate target nucleon for the \nuDIS cross-section which is measured by the neutrino telescopes.
The experiments measure the upward flux of neutrinos through the Earth, which is a mix of various elements. 
The NNPDF collaboration (amongst others) has calculated nuclear PDFs, going beyond the standard proton PDF, and in \cite{AbdulKhalek:2022fyi} they produced an ``Earth-average'' PDF, corresponding to nuclear mass number 31.
They found (see Figure 6.1 in that work) that the central values of the SM cross-section differ by less than \qty{5}{\percent} between a proton, oxygen and ``Earth average'' PDFs, and all three have comparable and overlapping \qty{10}{\percent} uncertainties.
A comparison of proton to isoscalar targets was done in \cite{Xie:2023suk} (see FIG.\ 16), again finding very similar cross-section results.%
\footnote{Further comparisons between deuterium, iron, or lead targets (Figure 5.5 of \cite{Candido:2023utz}) or nuclear targets with atomic mass numbers in the range A=\numrange{27}{33} (Section 5 of \cite{Garcia:2020jwr}) all show agreement within the roughly \qty{10}{\percent} error bands.}

We therefore conclude that at the current time each choice of target gives compatibly and equally valid results, and for simplicity we calculate the proton scattering cross-section, specifically using the ``\texttt{NNPDF40{\textunderscore}nnlo{\textunderscore}as{\textunderscore}01180}'' PDF set \cite{NNPDF:2021njg} accessed through the LHAPDF interface \cite{Buckley:2014ana}.
When we later compare the \nuDIS bounds to those from other methods, we neglect this theoretical uncertainty as we expect continued progress in PDFs will reduce it, and this gives us an idea of the future potential discovery power.

\section{Other constraints}
\label{sec:other_constraints}

\subsection{Direct LHC searches}
\label{sub:direct_lhc}

\subsubsection{$R_2$}

With our coupling structure, the $R_2$ LQ components each have two decay modes: $R_2^{5/3} \to c + \tau^+$ or $t + \tau^+$, and $R_2^{2/3} \to c \bar{\nu}_\tau$ or $b + \tau^+$, whose branching fractions are controlled in both cases by the relative size of the two couplings $g_L^{c\tau}$ and $g_R^{b\tau}$.
(Here again we neglect the CKM rotations, as this leads only to at most a few percent branching ratio of $R_2^{5/3} \to u \tau^+$.)
The branching ratios of these modes are:
\begin{align}
\text{Br} (R_2^{2/3} \to c \bar{\nu}_\tau) = \text{Br} (R_2^{5/3} \to c \tau^+) &= \frac{|g_L^{c\tau}|^2}{|g_L^{c\tau}|^2 + |g_R^{b\tau}|^2}
\\
\text{Br} (R_2^{2/3} \to b \tau^+) = \text{Br} (R_2^{5/3} \to t \tau^+) &= \frac{|g_R^{c\tau}|^2}{|g_L^{c\tau}|^2 + |g_R^{b\tau}|^2} \,,
\end{align}
where we have neglected all quark masses, since the $R_2$ mass will end up being greater than \qty{1}{\TeV}.
At the LHC, the relevant searches are therefore $\text{LQ} \to j \tau, t \tau, b \tau$ and $j \nu$, where $j = u,d,s,c$ represents a light quark jet.
Currently, the strongest searches for pair production of LQs with these decay modes comes from CMS for $j \nu$ \cite{CMS:2019ybf} and ATLAS for $t \tau$ \cite{ATLAS:2021oiz}, $b \tau$ \cite{ATLAS:2023uox}, and $j \tau$ \cite{ATLAS:2023kek}.
In \cite{ATLAS:2023uox} they directly provide limits on the branching ratio as a function of the LQ mass, while for the others we extract such a limit by comparing the experimental limit on $\sigma \times \text{Br}^2$ to the theoretical pair production cross-section.
\begin{figure}
\includegraphics[width=0.8\textwidth]{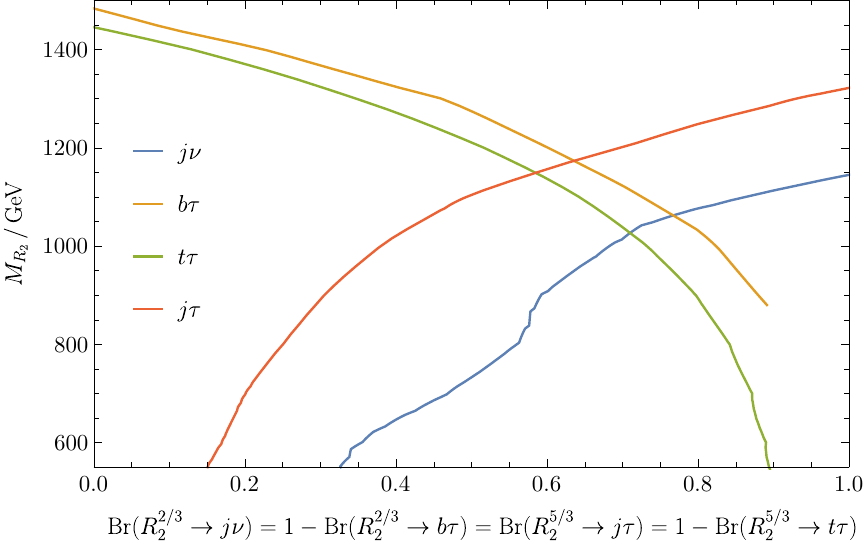}
\caption{Limits on the $R_2$ LQ from pair production at the LHC}
\label{fig:R2_LHC_pair_production}
\end{figure}
The combination of the LHC data is shown in \cref{fig:R2_LHC_pair_production}, and we find that the lowest allowed LQ mass is currently \qty{1175}{\GeV}, when the $R_2$ components decay \qty{65}{\percent} of the time to light jets + $\nu$ / $\tau$ respectively, which can be achieved by fixing $|g_R^{b\tau}| \approx 0.75 |g_L^{c\tau}|$.
As we will see, this relationship is compatible with a resolution of the $R_{D^{(*)}}$ anomalies, and so we adopt this ratio going forward.

\subsubsection{$S_1$}
\begin{figure}
\includegraphics[width=0.8\textwidth]{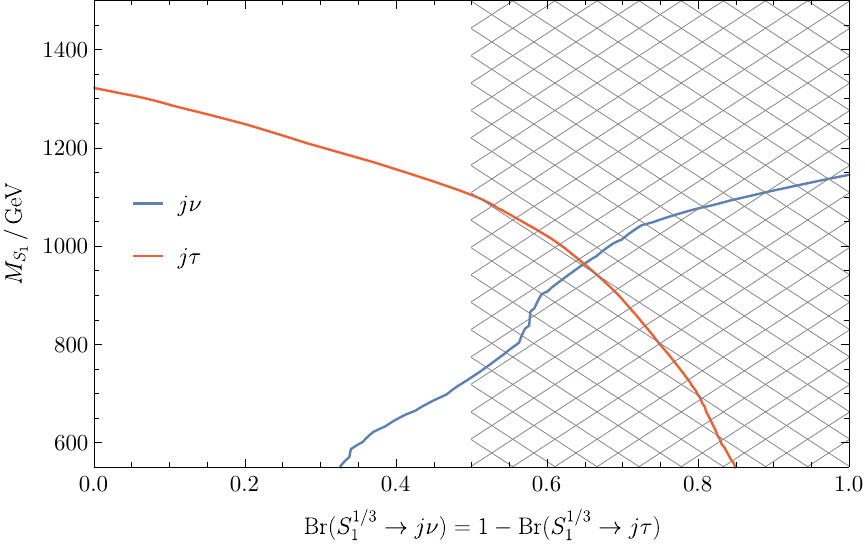}
\caption{Limits on the $S_1$ LQ from pair production at the LHC. The grey hatched is region inaccessible due to our coupling structure.}
\label{fig:S1_LHC_pair_production}
\end{figure}

We can conduct the same exercise for the $S_1$ LQ, whose single component has two decay modes $S_1^{1/3} \to \bar{s} + \bar{\nu}_\tau$ or $\bar{c} + \tau^+$ with branching fractions%
\footnote{Again neglecting CKM rotations and quark masses.}
\begin{equation}
\text{Br} (S_1^{1/3} \to \bar{s} \bar{\nu}_\tau) = \frac{|y_{LL}^{s\tau}|^2}{2|y_{LL}^{s\tau}|^2 + |y_{RR}^{c\tau}|^2} \,,
\quad
\text{Br} (S_1^{1/3} \to \bar{c} \tau^+) = \frac{|y_{LL}^{s\tau}|^2 + |y_{RR}^{c\tau}|^2}{2|y_{LL}^{s\tau}|^2 + |y_{RR}^{c\tau}|^2} \,.
\end{equation}
The relevant LHC searches here are therefore just $\text{LQ} \to j \nu$ or $j \tau$, for which we again use the CMS and ATLAS searches respectively, which we show in \cref{fig:S1_LHC_pair_production}.
For arbitrary couplings, the mass limit would be \qty{965}{\GeV}, however due to our specific coupling structure the branching ratio of $S_1 \to \bar{s} \bar{\nu}$ is always $\leq 0.5$ and so the LHC limits the mass to be at least greater than \qty{1100}{\GeV}.
If instead we fix $y_{LL}^{s\tau} = y_{RR}^{c\tau}$ in order to make a comparison of our neutrino bounds to those in \cite{Huang:2021mki}, the LQ mass must be \qty{1190}{\GeV} or higher.

We note that the recent ATLAS search for $j \tau$ final states \cite{ATLAS:2023kek} is the first explicit search for this decay mode%
\footnote{CMS searched for $b \tau$ final states in \cite{CMS:2018iye}, but did not make use of jet flavour tagging and so their results are actually also applicable to this case.}
and closes off the final avenue for light LQs, which was exploited in \cite{Huang:2021mki}.

\subsection{High $p_T$ Drell-Yan}
\label{sub:highPT}

LQs can also affect collider processes in an indirect way, through changes to the high momentum tails of Drell-Yan processes ($pp \to \ell \ell, \ell \nu$) (and this is known to be particularly important for explanations of the $R_{D^{(*)}}$ anomalies \cite{Faroughy:2016osc,Greljo:2018tzh,Mandal:2018kau,Aydemir:2019ynb}).
We use the tool \HighPT \cite{Allwicher:2022mcg,Allwicher:2022gkm} to evaluate the limits from di-tau \cite{ATLAS:2020zms} and mono-tau \cite{ATLAS:2021bjk} measurements.
In \HighPT, a full calculation for a small number of fixed LQ masses is provided, as well as results using the EFT framework of SMEFT operators%
\footnote{The tree-level matching of our LQs onto the SMEFT is given in \cref{app:smeft_matching}.}
that can be applied for arbitrary SMEFT Wilson coefficients.
We find that in each case the EFT bound is strictly stronger than the result using the full model dependence (see \cref{app:highpt_eft_vs_full} for details).
However the full limit scales roughly linearly for the fixed LQ masses provided, and therefore we show both a linear extrapolation of the full theory bound as well as the EFT constraint in our figures in \cref{sec:combination}.
Continuing our earlier discussion about the different CKM basis choices, we have checked and found that the Drell-Yan bounds are only very weakly sensitive to this choice.

\subsection{Flavour}
\label{sub:flavour}

\subsubsection{\texorpdfstring{$b \to c \ell \nu$}{b -> c l nu}}
\label{sub:bclnu}

For many years now, the experimental measurements of lepton flavour universality in $b \to c \ell \nu$ decays have been found to disagree with the SM (often collectively referred to as just the $R_{D^{(*)}}$ anomalies).
The latest HFLAV combination for $R_D$ and $R_{D^*}$, defined as 
\begin{equation}
R_{D^{(*)}} = \frac{\text{Br}(B \to D^{(*)} \tau \nu)}{\text{Br}(B \to D^{(*)} \ell \nu)} \quad (\ell = e, \mu) \,,
\end{equation}
is found to disagree with the SM prediction at just above \qty{3}{\sigma} \cite{HeavyFlavorAveragingGroup:2022wzx,HFLAV:Winter2023}.
It has long been known that the $R_2$ LQ can explain the observed anomalies in $b \to c \ell \nu$ transitions, as long as the low energy effective coupling of the $(\bar c b)(\bar \tau \nu)$ effective operator is large and mostly imaginary (e.g.\ see \cite{Tanaka:2012nw}).
This can simply be achieved by fixing our $g_L^{c\tau}$ coupling to be purely real, and $g_R^{b \tau}$ to be purely imaginary, such that their product is purely imaginary as well.

\begin{figure}
\includegraphics[width=0.8\textwidth]{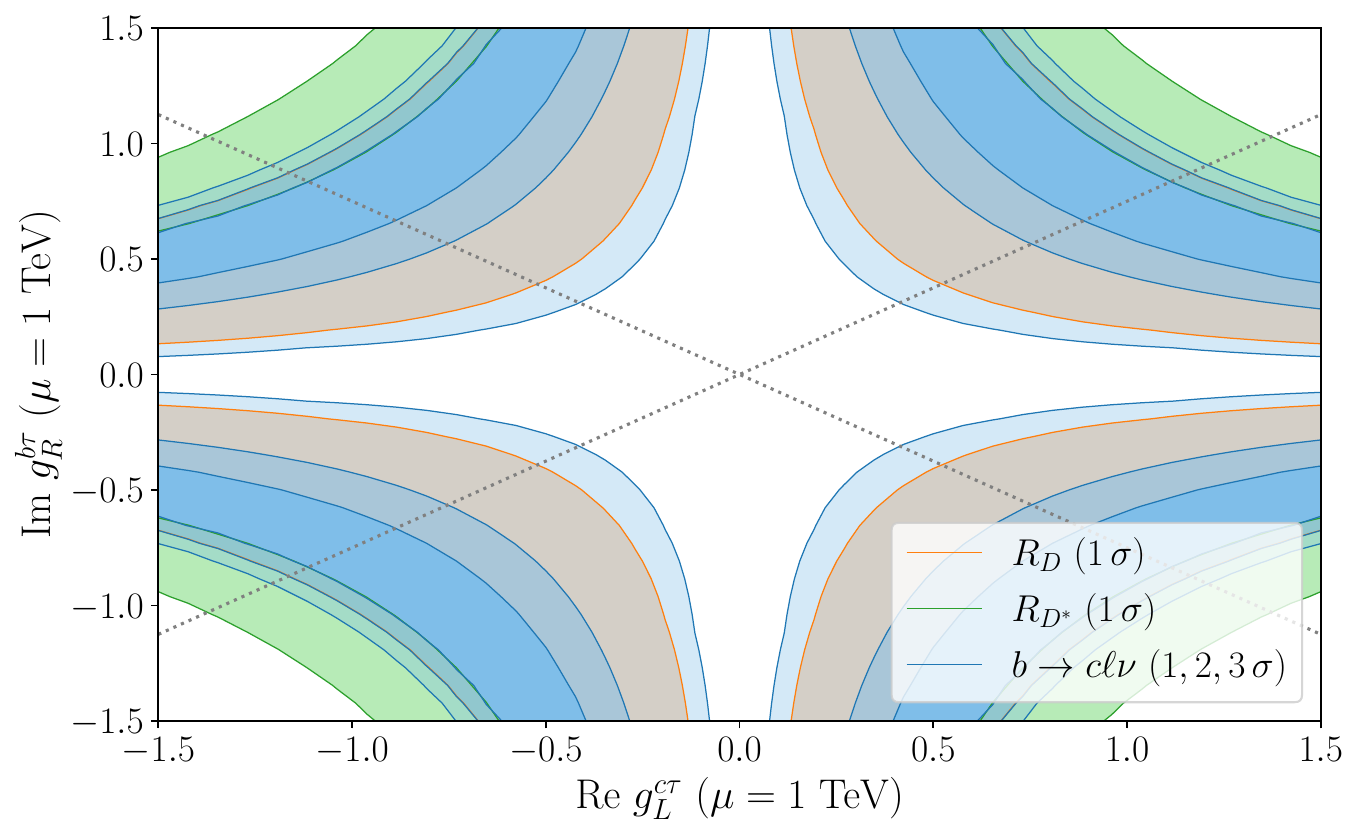}
\caption{Fit to $b \to c \ell \nu$ data for a \qty{1}{\TeV} $R_2$ LQ, along with grey dotted lines indicating our coupling ratio ansatz to minimise the LHC pair production bounds.}
\label{fig:R2_RD_fit}
\end{figure}

Using \smelli \cite{Aebischer:2018iyb,Stangl:2020lbh}%
\footnote{Which uses \flavio \cite{Straub:2018kue} for observable calculations and \wilson \cite{Aebischer:2018bkb} for RG running.}
we update the global fit to make use of all the latest data (in particular the recent updates from LHCb \cite{LHCb:2023cjr,LHCb:2023zxo}), and we show in \cref{fig:R2_RD_fit} the favoured regions from $R_D$, $R_{D^*}$, and all $b \to c \ell \nu$ observables.
The best fit region can be simply written as
\begin{equation}
\frac{0.12, 0.43, 0.59}{\unit{\TeV^2}} \lesssim \frac{|g_L^{c \tau} g_R^{b\tau}|}{M_{R_2}^2} \lesssim \frac{0.92, 1.0, 1.1}{\unit{\TeV^2}}
\end{equation}
for the 1, 2 and \qty{3}{\sigma} regions respectively, and the best fit point has a pull of \qty{3.1}{\sigma} relative to the SM, and thus our minimal coupling structure can explain the anomaly as we wanted.
Furthermore, the grey lines denote the relation $|g_R^{b\tau}| \approx 0.75 |g_L^{c\tau}|$, and thus we see that we can explain the $R_{D^{(*)}}$ anomalies with an $R_2$ LQ at the lightest LHC allowed mass of \qty{1175}{\GeV}.

\subsubsection{$S_1$ mediated leptonic decays}

Our $S_1$ LQ can mediate the decay $\tau \to K \nu$, as well as $D_s \to \tau \nu$, at tree level, neither of which were considered in \cite{Huang:2021mki}.
Both BSM contributions interfere with the SM as they have the same $\gamma_\mu P_L \otimes \gamma^\mu P_L$ Dirac structure, and in particular measurements of the tau decay generates strong bounds, which we find (using \flavio and the latest PDG experimental average \cite{ParticleDataGroup:2020ssz}) to be 
\begin{equation}
\frac{|y_{LL}^{c\tau}|}{M_{S_1}} < \frac{0.84}{\unit{\TeV}}
\end{equation}
at \qty{2}{\sigma}.

\subsection{EDM}
\label{sub:EDM}

Since in the $R_2$ model we have a large coupling to both left and right handed quarks, and one is imaginary, electric dipole moments (EDMs) are induced.
In \cite{Dekens:2018bci}, 
all low energy couplings relevant for the EDMs were calculated by matching the LQs with the SMEFT at \qty{1}{\TeV} and evolving the Wilson coefficients down to the hadronic scale.
With our coupling structure, 
the most relevant result is the charm EDM $d_c$ (see Table 1 of \cite{Dekens:2018bci}), given by
\begin{equation}
d_c \simeq 0.1 \times e \, Q_c \, m_c \frac{\Im(\V*{cb}\, g_R^{b\tau*} g_L^{c\tau})}{M_{R_2}^2} 
\end{equation}
where the LQ couplings are evaluated at $\mu = \qty{1}{\TeV}$.

Currently, strong experimental limits are set on both neutron and mercury EDMs.
The \qty{2}{\sigma} upper limit is $|d_n| < \qty{2.2 e-26}{e.\cm}$ \cite{Abel:2020pzs} and $|d_\text{Hg}| < \qty{7.9 e-30}{e.\cm}$ \cite{Graner:2016ses}.
In our case, neutron EDM is generated mostly from the charm EDM via $d_n \simeq g_T^c d_c$, where $g_T^c$ is the charm tensor charge, and the mercury EDM is given by
\begin{equation}
d_\text{Hg} \simeq -\num{2.1\pm 0.5 e-4} \left[ (1.9 \pm 0.1) d_n + (0.20 \pm 0.06) d_p \right] 
\end{equation}
where $d_p \simeq d_n$.

Given the recently calculated charm tensor charge $g_T^c = \num{-2.4(1.6) e-4}$ \cite{Alexandrou:2019brg},
we find 
\begin{equation}
\frac{\left| \Im (g_L^{c\tau} g_R^{b\tau *}) \right|}{M_{R_2}^2} \leq \frac{1.3}{\unit{\TeV^2}}
\end{equation}
from the neutron bound, or 
\begin{equation}
\frac{\left| \Im (g_L^{c\tau} g_R^{b\tau *}) \right|}{M_{R_2}^2} \leq \frac{1.1}{\unit{\TeV^2}}
\end{equation}
from mercury, where we have used the central value of the charm tensor charge, and so we note that this bounds could easily be much weaker or stronger, depending on a future precise lattice determination of this quantity.

\subsection{EWPO}
\label{sub:ewpo}

At 1-loop, our LQs modify the $Z \to \tau \tau$ decay, with potentially large contributions coming from loops with top quarks.
The most accurate calculation of these LQs contributions (including finite terms and the corrections due to the external momenta of the electroweak bosons for the first time) was performed in \cite{Arnan:2019olv} (and later confirmed in \cite{Crivellin:2020mjs}), whose results we use.

The $R_2$ LQ has a direct (i.e.\ non-CKM suppressed) coupling between tau leptons and top quarks through the $g_R^{b\tau}$ coupling, and so non-zero values are strongly constrained by data.
With our assumed ratio of $g_R^{b\tau} = 0.75i \times g_L^{c\tau}$, we find a good approximation to the \qty{2}{\sigma} bound in the region we consider to be
\begin{equation}
g_{L}^{c\tau} \lesssim 0.9 \left(\frac{M_{R_2}}{\unit{\TeV}}\right) + 0.4 \,.
\end{equation}

For the $S_1$ however, the top-tau vertex is suppressed by a factor of \V{ts}, and we find much weaker bounds apply. 
Again a good approximation to the \qty{2}{\sigma} bound can be simply written as 
\begin{equation}
y_{LL}^{s\tau} \lesssim 3.3 \left(\frac{M_{S_1}}{\unit{\TeV}}\right) + 0.55 \,.
\end{equation}

\section{Probing the parameter space}
\label{sec:combination}

As we have seen from \cref{fig:nuDIS_xsec} the LQ enhancement of the DIS cross-section grows with energy, so we optimistically use the $E_\nu = \qty{1e10}{\GeV}$ enhancement to show the potential parameter space probed.
Similarly, we ignore any theoretical uncertainties from the PDFs (as discussed in \cref{sub:nu_pdf_errors}) to give the best possible case for the neutrino experiments and highlight the fundamental future potential reach (noting that in this case, to make use of a neutrino measurement at the highest precision would require improvements in the PDF uncertainties beyond what is currently available).
As discussed earlier in \cref{sub:nu_searches}, we will use the two proposed scenarios from \cite{Esteban:2022uuw}, of being able to measure the \nuDIS cross-section to either ${}^{+\qty{65}{\percent}}_{-\qty{30}{\percent}}$ with a small number of UHE neutrino events or $\pm\qty{15}{\percent}$ with larger statistics, as the basis for our parameter space bounds.

\subsection{$R_2$}

\begin{figure}
\includegraphics[width=\textwidth]{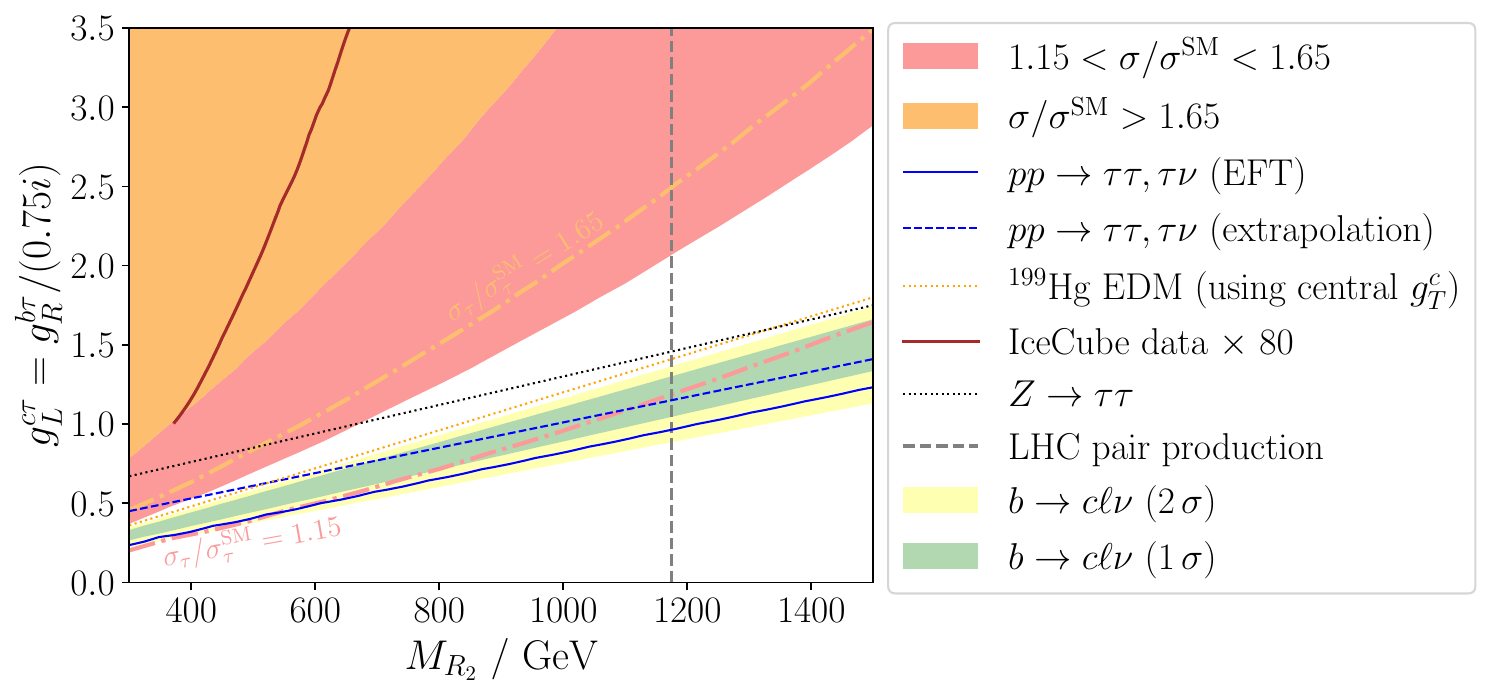}
\caption{Constraints on the parameter space of an $R_2$ LQ that can explain the $R_{D^{(*)}}$ anomalies. The dot-dash lines indicate the bound for a tau flavour only measurement, i.e. ($\sigma_\tau/\sigma_\tau^\text{SM}$). All bounds are at \qty{2}{\sigma} unless otherwise specified. The ``IceCube data $\times$ 80'' bound is taken from Figure 14 of \cite{Becirevic:2018uab}.}
\label{fig:R2_combined_constraints}
\end{figure}

In \cref{fig:R2_combined_constraints} we show our bounds from all the searches already considered, along with the limit derived in \cite{Becirevic:2018uab} from a future IceCube data sample of 80 times the size available at the time.
The first result we note is that our neutrino bounds are stronger than those from IceCube, even in the conservative case of \qty{65}{\percent} precision on a flavour averaged cross-section from a small number of observed UHE neutrino events, while the optimistic future reach goes far beyond that, probing LQ masses of over a \unit{\TeV} (albeit at large couplings).
This result can be understood by considering that the IceCube data uses neutrinos at energies in the range \qtyrange{1e5}{1e7}{\GeV}, where we can see from \cref{fig:nuDIS_xsec} that the LQ enhancements are much smaller than at the \qty{1e10}{\GeV} energies we consider.
However, we find that the latest LHC pair production searches and the Drell-Yan bounds (as well as $Z \to \tau \tau$ and potentially the mercury EDM measurements) are better probes of the parameter space for an $R_2$ LQ that explains the $R_{D^{(*)}}$ anomalies, even in the best case scenario where our tau specific NP model is probed by a tau only measurement of the \nuDIS cross-section (and we expect that the collider constraints will only get stronger in the future, see for example studies looking at prospects at the HL-LHC \cite{Iguro:2020keo,Bhaskar:2021gsy,Endo:2021lhi}, or beyond to the FCC-ee or CEPC experiments \cite{Fedele:2023gyi,Ho:2022ipo}).

As discussed earlier, the $Z \to \tau \tau$ measurements strongly restrict non-zero values of $g_R^{b\tau}$. 
We therefore show in \cref{app:R2_gR0} an equivalent plot to \cref{fig:R2_combined_constraints} with this coupling set to zero for comparison, which has a minimal effect on the neutrino experiment bounds, but removes the $Z$ bounds, as well as those from the EDMs and $R_{D^{(*)}}$.

\subsection{$S_1$}

\begin{figure}
\includegraphics[width=\textwidth]{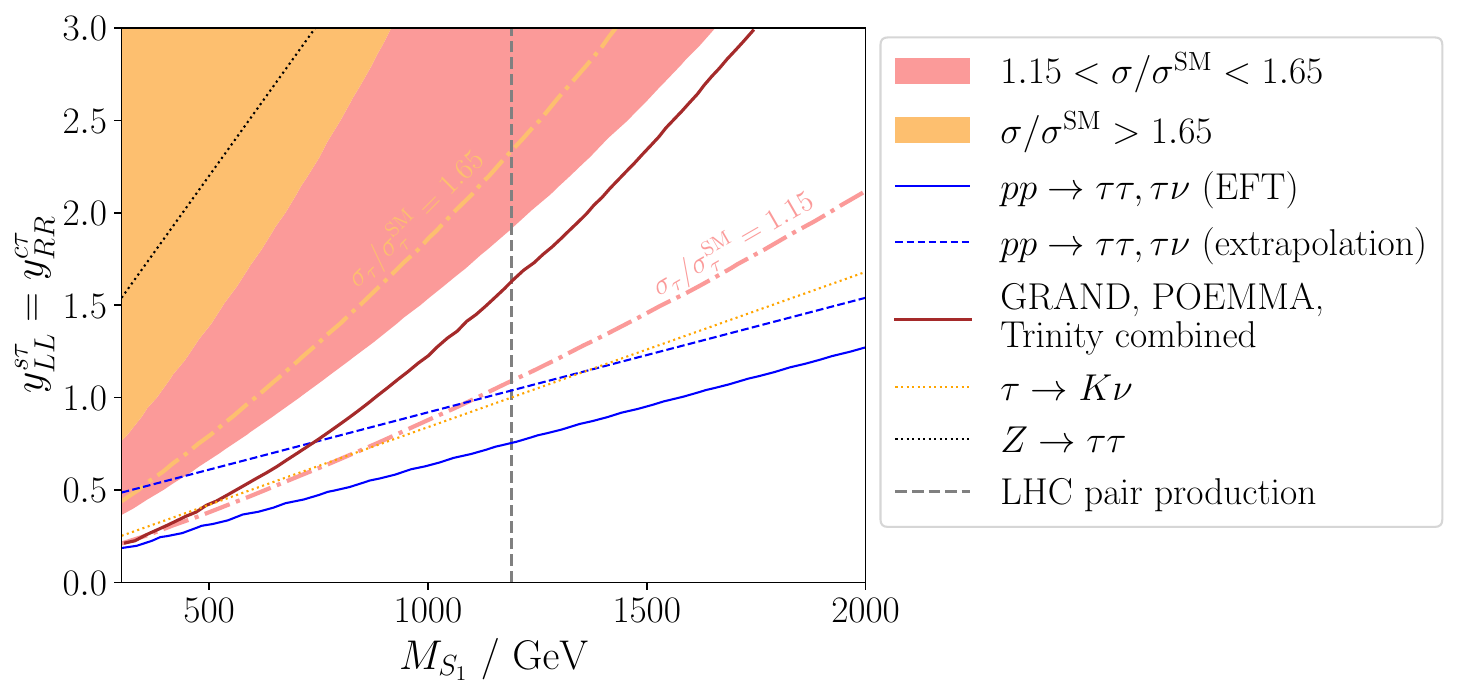}
\caption{Constraints on the parameter space of an $S_1$ LQ. The dot-dash lines indicate the bound for a tau flavour only measurement, i.e. ($\sigma_\tau/\sigma_\tau^\text{SM}$). All bounds are at \qty{2}{\sigma} unless otherwise specified. The ``GRAND, POEMMA, Trinity combined'' bound is taken from Figure 10 of \cite{Huang:2021mki}.}
\label{fig:S1_combined_constraints}
\end{figure}

Our other case study is shown in \cref{fig:S1_combined_constraints}.
Here we see that our neutrino bound is weaker than the combined bound from GRAND, POEMMA and Trinity that was found in \cite{Huang:2021mki} if we assume a measurement is made of the flavour averaged cross-section, while if the best case scenario is realised of a tau specific cross-section measurement, the sensitivity reported by \cite{Esteban:2022uuw} in an optimistic scenario of around 100 neutrino events exceeds their result.
This is despite the assumptions in \cite{Huang:2021mki} of a larger ($>100$) number of detection events in their analysis, as well as the fact that (as discussed in \cref{sub:nu_searches}) they perform a more detailed analysis of the experiments and models and use more information from the BSM neutrino spectrum.
On the other hand, the method of \cite{Esteban:2022uuw} avoids the uncertainty around the incoming neutrino flux which has to be marginalised over in \cite{Huang:2021mki}, and our bounds are absolute best case scenarios since we neglect PDF uncertainties as already discussed.
However, we find that here the new searches from ATLAS for the $\text{LQ} \to j \tau$ have massively increased the mass limit for an $S_1$ with our chosen couplings (from \num{400} to \qty{1190}{\GeV}), and combined with better Drell-Yan data, we are able to rule out the low mass parameter space that was previously only probed by the neutrino experiments.
In addition, the flavour bounds we find from the $\tau \to K \nu$ decay, which were not considered in \cite{Huang:2021mki}, are similarly strong (although as noted earlier those are only valid in the CKM basis we have chosen).

\section{Conclusions}
\label{sec:conclusions}

Next generation UHE neutrino experiments have the potential to provide new insights into interactions of neutrinos with matter at the ultrahigh energies.
In this paper, focusing on two motivated LQs, $R_2$ and $S_1$, we have explored the possibility to probe at the neutrino experiments the enhancement of the neutrino-nucleon scattering which is an intrinsic feature in any LQ model.
Building on the projected sensitivities estimated in \cite{Esteban:2022uuw} (which agree with other recent projections in \cite{Denton:2020jft} and \cite{Valera:2022ylt}), we have made a full calculation of the \nuDIS cross-section for our two chosen LQ models, and discussed how improvement to PDF and related uncertainties would be needed to fully unlock the potential of UHE neutrino data.
We then studied the sensitivity of upcoming neutrino experiments, as well as current LHC, flavour and EWPO constraints, performing an analysis of the most recent data for each, and found that recent LHC searches for LQ pair-production have closed off a gap in parameter space that allowed light LQs. 
For heavier LQs, the latest LHC measurements of high-$p_T$ tails for $pp\to\tau\tau, \tau\nu$ processes already exclude parameter space that will be within reach of the future neutrino experiments, even in the best case scenario.
It is therefore unlikely that for LQ searches the future precision expected from neutrino telescopes is a match for the power of the LHC.

\section*{Acknowledgements}
All authors are very grateful to Ivan Esteban for useful help and discussions during the early stages of this project, and to the authors of \cite{Huang:2021mki} for their help with the results of their paper.
M.K.\ thanks James Whitehead and Juan M. Cruz-Martinez for useful discussions on PDFs.
M.K.\ and K.W.\ thank Federico Mescia for help with the results of \cite{Arnan:2019olv}.
M.K.\ and S.O.\ acknowledge support from a Maria Zambrano fellowship, and K.W.\ is supported by the China Scholarship Council (CSC).
All authors acknowledge support from the State Agency for Research of the Spanish Ministry of Science and Innovation through the ``Unit of Excellence María de Maeztu 2020-2023'' award to the Institute of Cosmos Sciences (CEX2019-000918-M), and from PID2019-105614GB-C21 and 2017-SGR-929 grants.

\appendix

\section{$R_2$ with \texorpdfstring{$g_R^{b\tau} = 0$}{gRbtau = 0}}
\label{app:R2_gR0}

\begin{figure}
\includegraphics[width=\textwidth]{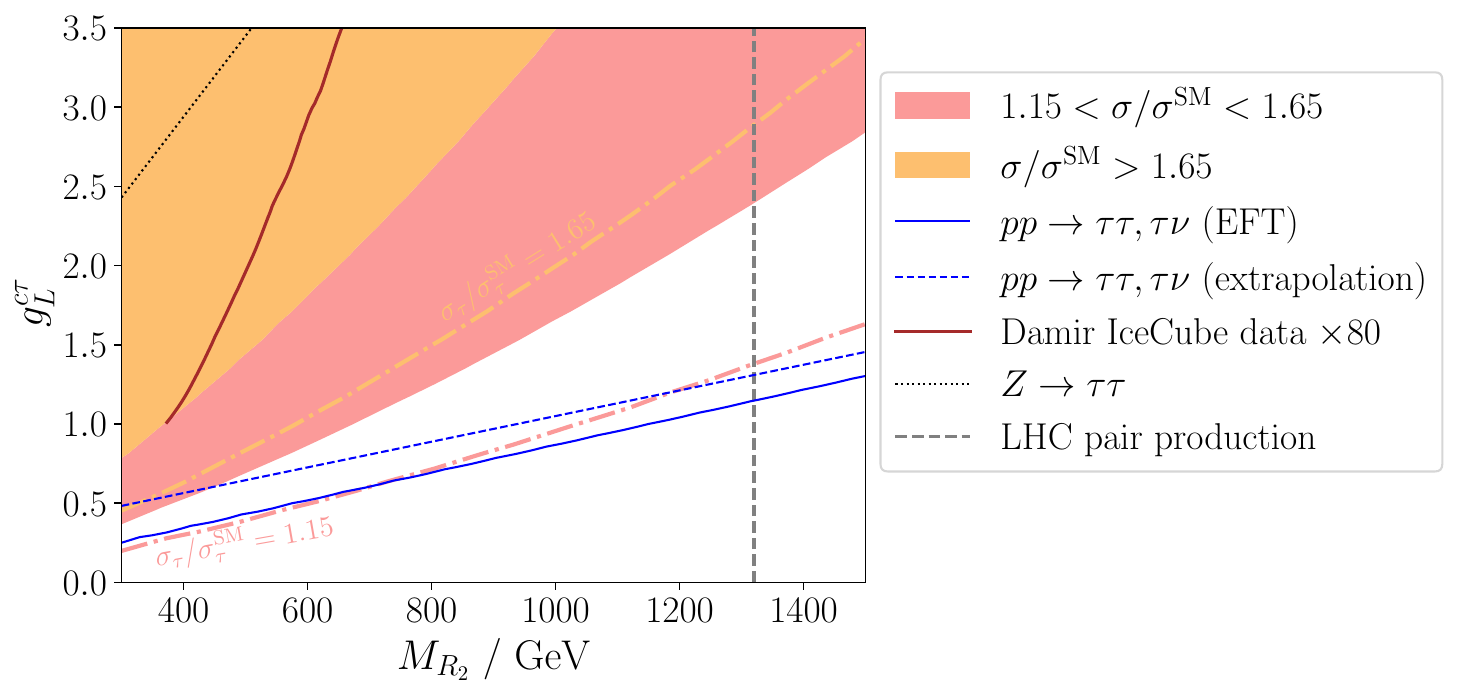}
\caption{Constraints on the parameter space of an $R_2$ LQ with $g_R^{b\tau} = 0$. The dot-dash lines indicate the bound for a tau flavour only measurement, i.e. ($\sigma_\tau/\sigma_\tau^\text{SM}$). All bounds are at \qty{2}{\sigma} unless otherwise specified. The ``IceCube data $\times$ 80'' bound is taken from Figure 14 of \cite{Becirevic:2018uab}.}
\label{fig:R2_combined_constraints_gR0}
\end{figure}

In the main text, we fixed a non-zero value of the $R_2$ coupling $g_R^{b\tau}$ in order to allow an explanation of the $R_{D^{(*)}}$ anomalies.
If we give up this property and set $g_R^{b\tau} = 0$, we find much weaker constraints from $Z \to \tau \tau$ of $g_{L}^{c\tau} \lesssim 5.1 \left(M_{R_2} / \unit{\TeV}\right) + 0.9$ (since the top loop contribution from $g_L$ is CKM suppressed), and that the EDM bounds vanishes entirely.
Meanwhile the Drell-Yan searches and the neutrino scattering cross-section turn out to be almost independent of the value of this coupling.
In \cref{fig:R2_combined_constraints_gR0} we show an alternative version of \cref{fig:R2_combined_constraints} for comparison.
We see that LHC data, from both direct searches and Drell-Yan studies, remain much more constraining than the neutrino experiments.

\section{SMEFT matching}
\label{app:smeft_matching}

Here we state the tree-level SMEFT matching coefficients, as used in the Drell-Yan and flavour constraints.
\paragraph{$R_2$}
\begin{equation}
\begin{aligned}
\left[C_{lequ}^{(1)}\right]_{3332} &= -\frac{(g_L^{c\tau})^* g_R^{b\tau}}{2 M_{R_2}^2} \,,
&
\left[C_{lequ}^{(3)}\right]_{3332} &= -\frac{(g_L^{c\tau})^* g_R^{b\tau}}{8 M_{R_2}^2} \,,
\\
\left[C_{qe}\right]_{3333} &= -\frac{|g_R^{b\tau}|^2}{2 M_{R_2}^2} \,,
&
\left[C_{lu}\right]_{3322} &= -\frac{|g_L^{c\tau}|^2}{2 M_{R_2}^2} \,.
\end{aligned}
\end{equation}

\paragraph{$S_1$}
\begin{equation}
\begin{aligned}
\left[C_{lq}^{(1)}\right]_{3322} &= \frac{|y_{LL}^{s\tau}|^2}{4 M_{S_1}^2} \,,
&
\left[C_{lq}^{(3)}\right]_{3322} &= -\frac{|y_{LL}^{s\tau}|^2}{4 M_{S_1}^2} \,,
\\
\left[C_{lequ}^{(1)}\right]_{3322} &= \frac{(y_{LL}^{s\tau})^* y_{RR}^{c\tau}}{2 M_{S_1}^2} \,,
&
\left[C_{lequ}^{(3)}\right]_{3322} &= -\frac{(y_{LL}^{s\tau})^* y_{RR}^{c\tau}}{8 M_{S_1}^2} \,,
\\
\left[C_{eu}\right]_{3322} &= \frac{|y_{RR}^{c\tau}|^2}{2 M_{S_1}^2} \,.
\end{aligned}
\end{equation}

\section{Neutrino DIS matrix elements}
\label{app:nuDIS_matrix_elements}
The remaining neutrino DIS matrix elements, beyond those already given in \cref{sub:nuDIS_xsec} and which have a minimal contribution to the total cross-section, are given below.

\paragraph{$R_2$}

\begin{align}
\langle |M(\nu_i u_j \to \nu_k u_l)|^2 \rangle &= 
\begin{aligned}[t]
&\frac{\hat{u}^2}{2} \frac{|g_L^{jk}|^2 |g_L^{li}|^2}{(\hat{u} - M_{R_2}^2)^2}
\\
&+ \delta_{ik} \delta_{jl} \left( 
-\frac{8 \alpha \pi}{3 c_W^2} \frac{\hat{u}^2}{(\hat{u} - M_{R_2}^2)(\hat{t}-M_Z^2)} \Re(g_L^{li} g_L^{jk*})
\right.
\\
 &\qquad \left.
 + \frac{2 \alpha^2 \pi^2 (16 \hat{u}^2 s_W^4 + (3-4 s_W^2)^2 \hat{s}^2)}{9 c_W^4 s_W^4 (\hat{t} - M_Z^2)^2}
 \right)	
\end{aligned}
\\
\langle |\mathcal{M}(\nu_i d_j \to \nu_k d_l)|^2 \rangle &=
\begin{aligned}[t]
\frac{2 \alpha^2 \pi^2 ((3-2 s_W^2)^2 \hat{s}^2 + 4 \hat{u}^2 s_W^4)}{9 c_W^4 s_W^4 (\hat{t} - M_Z^2)^2} \delta_{ik} \delta_{jl}
\end{aligned}
\\
\langle |\mathcal{M}(\nu_i d_j \to \ell_k u_l)|^2 \rangle &= 
\begin{aligned}[t]
&\frac{\hat{u}^2}{2} \frac{|g_L^{li}|^2 |g_R^{kj}|^2}{(\hat{u} - M_{R_2}^2)^2}
\\
&+ \delta_{ik} \delta_{jl} \left( + \frac{8 \alpha^2 \pi^2 \hat{s}^2}{s_W^4 (\hat{t} - M_W^2)^2}
 \right)
\end{aligned}
\\
\langle |\mathcal{M}(\nu_i \bar{d}_j \to \nu_k \bar{d}_l)|^2  \rangle &=
\begin{aligned}
\frac{2 \alpha^2 \pi^2 (4 \hat{s}^2 s_W^4 + (3-2 s_W^2)^2 \hat{u}^2)}{9 c_W^4 s_W^4 (\hat{t} - M_Z^2)^2} \delta_{ik} \delta_{jl}
\end{aligned}
\end{align}

\paragraph{$S_1$}

\begin{align}
\langle |M(\nu_i \bar{u}_j \to \nu_k \bar{u}_l)|^2 \rangle &= 
\begin{aligned}[t]
\frac{2 \alpha^2 \pi^2}{9 c_W^4 s_W^4} \frac{16 \hat{s}^2 s_W^4 + \hat{u}^2 (-3 c_W^2 + s_W^2)^2}{(\hat{t} - M_Z^2)^2} \delta_{ik} \delta_{jl}
\end{aligned}
\\
\langle |M(\nu_i u_j \to \nu_k u_l)|^2 \rangle &=
\begin{aligned}
\frac{2 \alpha^2 \pi^2}{9 c_W^2 s_W^2}	\frac{\hat{s}^2 (-3 c_W^2 + s_W^2)^2 + 16 \hat{u}^2 s_W^4}{(\hat{t} - M_Z^2)^2} \delta_{ik} \delta_{jl}
\end{aligned}
\\
\langle |M(\nu_i \bar{u}_j \to \ell_k \bar{d}_l)|^2 \rangle &=
\begin{aligned}[t]
&\frac{\hat{u}^2 |y_{LL}^{li}|^2 (|y_{LL}^{jk}|^2 + |y_{RR}^{jk}|^2)}{2 (\hat{u} - M_{S_1}^2)^2}
\\
&+ \delta_{ik} \delta_{jl} \left( \frac{4 \alpha \pi}{s_W^2} \frac{\hat{u}^2}{(\hat{u} - M_{S_1}^2)^2 (\hat{t} - M_W^2)} \Re(y_{LL}^{li *} y_{LL}^{jk}) \right.
\\
&\qquad \left.  \frac{8 \alpha^2 \pi^2}{s_W^4} \frac{\hat{u}^2}{(\hat{t} - M_W^2)^2}\right)
\end{aligned}
\\
\langle |M(\nu_i \bar{d}_j \to \nu_k \bar{d}_l)|^2 \rangle &=
\begin{aligned}[t]
&\frac{\hat{u}^2 |y_{LL}^{jk}|^2 |y_{LL}^{li}|^2}{2 (\hat{u} - M_{S_1}^2)^2} 
\\
&+ \delta_{ik} \delta_{jl} \left( - \frac{2 \alpha \pi}{3 c_W^2 s_W^2} \frac{\hat{u}^2 (-3 + 2 s_W^2)}{(\hat{u} - M_{S_1}^2) (\hat{t} - M_Z^2)} \Re( y_{LL}^{li*} y_{LL}^{jk}) \right.
\\
&\qquad \left. \frac{2 \alpha^2 \pi^2}{9 c_W^4 s_W^4} \frac{4 \hat{s}^2 s_W^4 + \hat{u}^2 (3-2s_W^2)^2}{(\hat{t} -  M_Z^2)^2} \right)
\end{aligned}
\end{align}

\section{Comparison of EFT and full Drell-Yan bounds}
\label{app:highpt_eft_vs_full}

\begin{figure}
\includegraphics[width=0.48\textwidth]{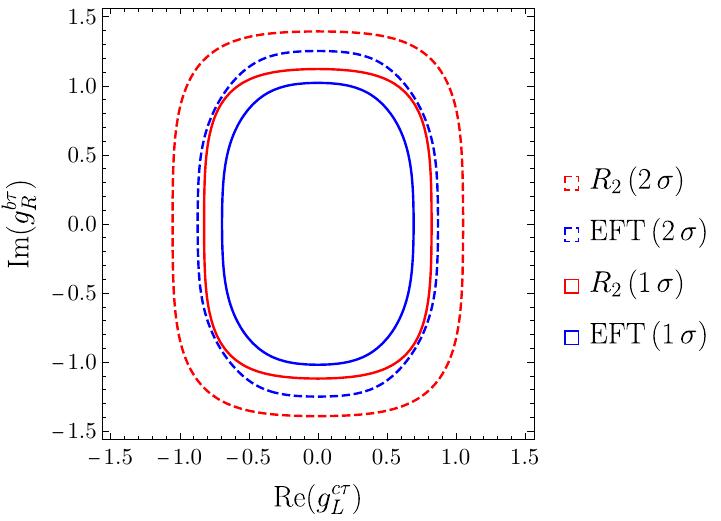}
\hfill
\includegraphics[width=0.48\textwidth]{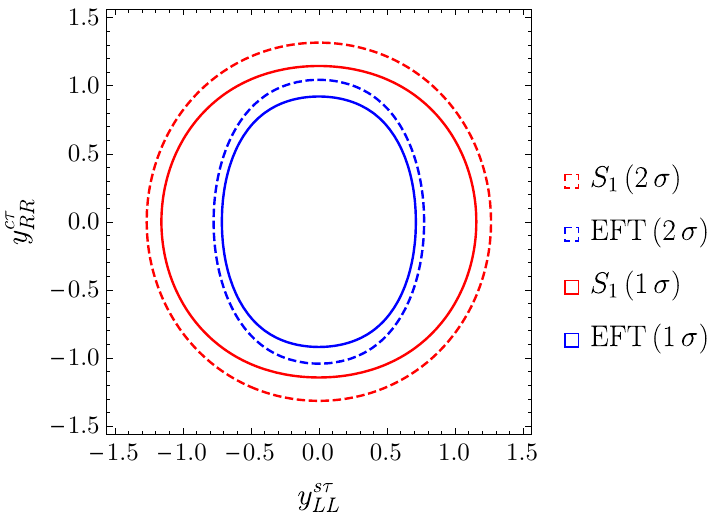}
\caption{A comparison of the full theory (red) and EFT (blue) bounds from Drell-Yan data using \HighPT, for our $R_2$ LQ (left) or $S_1$ LQ (right) at a mass of \qty{1}{\TeV}.}
\label{fig:LQ_hightpt_eft_vs_full}
\end{figure}
In \cref{fig:LQ_hightpt_eft_vs_full} we show the 1 and \qty{2}{\sigma} bounds (solid and dashed respectively) on our LQ models at masses of \qty{1}{\TeV}, as found using both the EFT (blue) and full theory (red) calculation mode of \HighPT.
We see that in both cases, for all values of the couplings, the EFT bound is stronger than the corresponding result using the full theory calculation.

\section{Neutrino regeneration}
\label{app:nu_regeneration}

The differential cross-section relates to the energy and direction of the neutrinos scattered inside the Earth, and hence to neutrino regeneration (i.e. neutrinos which interact but are subsequently detected).
In \cite{Esteban:2022uuw}, they studied regeneration in the SM and found it subleading, which meant they could place model-independent bounds on the total cross-section.

\begin{figure}
\includegraphics[width=\textwidth]{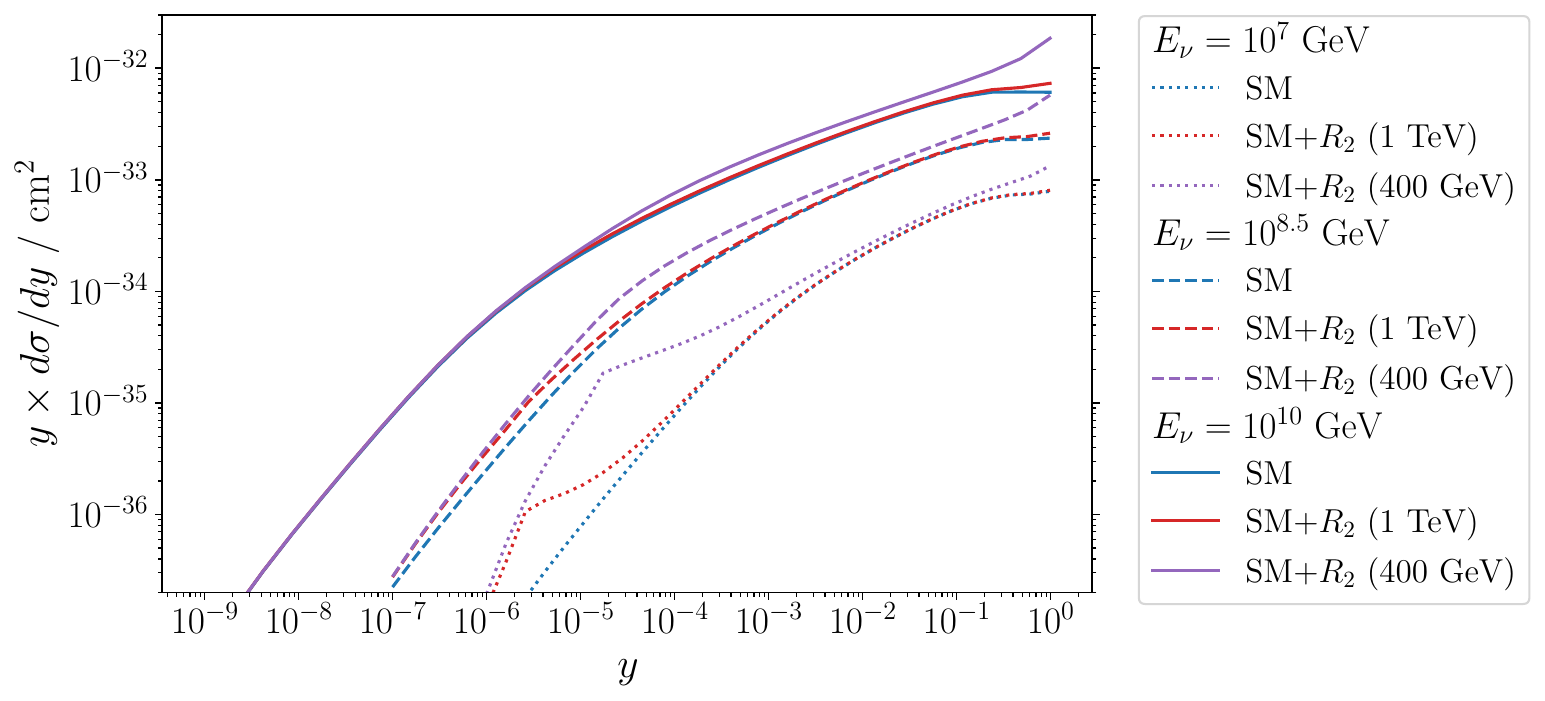}
\caption{Comparison of differential cross-section between the SM (blue) and our $R_2$ LQ scenario (where we have fixed $g_L^{c\tau} = g_R^{b\tau} / (0.75i) = 1$), at different neutrino energies (solid vs dashed vs dotted) and LQ masses (red and purple).}
\label{fig:dsigma_dy_comparision}
\end{figure}

In \cref{fig:dsigma_dy_comparision} we show the differential cross-section $d \sigma / dy$ for different neutrino energies and different masses of the $R_2$ LQ we study in this paper (extremely similar results can be found for the $S_1$ LQ).
As we can see, the differential cross-section is very similar to the SM spectrum for both LQ masses at high neutrino energies, and the significant deviation from the SM cross section shows up mostly in an inelastic scattering region, namely $y \sim 0.1 - 1$.
In particular, since we make use of the BSM enhancement at the highest energies ($E_\nu = \qty{1e10}{\GeV}$) since this is where it is largest, and our LHC analysis shows that our LQ candidates must have masses of at least \qty{1}{\TeV}, the most relevant comparison for our purposes is between the solid blue and red lines, which show almost total agreement.
This further justifies our application of the cross-section bounds from \cite{Esteban:2022uuw} to our LQ models.

\bibliographystyle{JHEP_MJKirk}
\bibliography{references}

\end{document}